\documentclass{elsart1p}
\usepackage{graphicx}
\usepackage{amssymb}

\newcommand{\sqrtsNN}{\mbox{$\sqrt{\mathrm{s}_{_{\mathrm{NN}}}}$} }
\newcommand{\vtwo}{$v_{2}$ }
\newcommand{\ptt}{$p_{T}$ }
\newcommand{\ks}{${K}^{0}_{S}$ }
\newcommand{\lam}{$\Lambda$ }

\def \GeVc {\mbox{$\mathrm{GeV}/c$}}
\def \GeVcsqr {\mbox{$\mathrm{GeV}/c^{2}$}}

\def \auau  {$\mathrm{Au + Au}$ }

\begin{document}
\begin{frontmatter}
% notes:
% use the corauthref command within \author for corresponding author
% footnotes;
\title{The elliptic flow of multi-strange hadrons in \sqrtsNN = 200 GeV
Au + Au collisions at STAR}
\thanks[label1]{The author was supported in part by the National Natural Science
Foundation of China under grant no. 10775058, 11075060 and self-determined research
funds of CCNU from the colleges'basic research and operation of MOE.}

\author{Shusu Shi$^{a, b}$ for the STAR Collaboration}

\address[a]{Institute of Particle Physics, Huazhong
Normal University, Wuhan, Hubei, 430079, China}
\address[b]{The Key Laboratory of Quark and Lepton Physics (Huazhong Normal
University), Ministry of Education, Wuhan, Hubei, 430079, China}

\begin{abstract}
The measurement of the elliptic flow, $v_{2}$,
provides a powerful tool for studying the properties of hot and
dense medium created in high-energy nuclear collisions.  We present the measurement of $v_{2}$ of
multi-strange hadrons ($\phi$, $\Xi$ and $\Omega$) in \sqrtsNN = 200 GeV Au + Au collisions at STAR.
In minimum-bias \auau collisions at $\sqrt{\mathrm{s}_{_{\mathrm{NN}}}}$ = 200 GeV, a significant amount
of elliptic flow, almost identical to other mesons and baryons, is
observed for multi-strange hadrons. Experimental observations of
$p_{T}$ dependence of $v_{2}$ of identified particles at RHIC
support partonic collectivity. We also discuss the possible breaking of mass ordering
of the $\phi$ mesons in the low \ptt region.

\end{abstract}

\end{frontmatter}

% main text
%\section{Introduction}
In the non-central nucleus nucleus collisions, the overlapping region of the reaction zone
is not spherical.  There is a short axis which is parallel to the impact parameter and a long axis
which is perpendicular to it. The initial geometrical anisotropy in the coordinated space will
be translated to the final anisotropy in the momentum space by the interactions of constituents.
This effect is characterized by the elliptic flow, $v_{2}$,
which is the second order harmonic
of the Fourier expansion of particle's azimuthal distribution
with respect to the reaction plane, defined as
\begin{equation} v_{2}=\langle\cos2(\phi-\Psi_{R})\rangle
\end{equation}where $\phi$ is azimuthal angle of an outgoing particle,
$\Psi_{R}$ is the azimuthal angle of the impact parameter, and
angular brackets denote an average over many particles and events.
As the system expands it becomes more spherical, thus the driving force quenches itself.
Therefore the elliptic flow is sensitive to the collision dynamics in the early stages.
It has been proved that the elliptic flow is one of the most sensitive
probes of the dynamics at the Relativistic Heavy
Ion Collider (RHIC)~\cite{flow1,flow2,flow3,flow4}, also see recent review
in~\cite{review1, review2, review3}.
However, early dynamic information might be masked by
later hadronic rescatterings. Multi-strange hadrons ($\phi$, $\Xi$
and $\Omega$) with their large mass and presumably small hadronic
cross sections~\cite{MSH_section0, MSH_section1} should be less sensitive to hadronic rescattering in
the later stage of the collisions and therefore a good probe of the
early stage of the collision.

%\section{Methods and Analysis}
In this paper, we present \vtwo measurements of multi-strange hadrons
by the STAR experiment from \sqrtsNN = 200 GeV \auau. Data
were taken from the seventh RHIC run in 2007.
About 63 million minimum bias events ($0-80\%$ most central) were analyzed.
STAR's Time
Projection Chamber (TPC)~\cite{STARtpc} is used as the
main detector for particle identification (PID) and event plane determination. The
centrality was determined by the number of tracks from the pseudorapidity region
$|\eta|\le 0.5$.
The PID is achieved via $dE/dx$ in TPC and topologically reconstructed hadrons:
\ks $\rightarrow \pi^{+} + \pi^{-}$, $\phi \rightarrow K^{+} +
K^{-}$, \lam $\rightarrow p + \pi^{-}$ ($\overline{\Lambda}
\rightarrow \overline{p} + \pi^{+}$),
$\Xi^{-} \rightarrow$ \lam $+\ \pi^{-}$ ($\overline{\Xi}^{+}
\rightarrow$ $\overline{\Lambda}$+\ $\pi^{+}$) and $\Omega^{-}
\rightarrow$ \lam $+\ K^{-}$ ($\overline{\Omega}^{+} \rightarrow$
$\overline{\Lambda}$+\ $K^{+}$). The detailed description of the
procedure can be found in Refs.~\cite{klv2_130GeV,starklv2, XiOmega}.
The event plane method~\cite{v2Methods1, v2Methods2} is used for the $v_{2}$ measurement.

%\section{Results and Discussions}

\begin{figure*}[ht]
\centering \hskip -.0cm \vskip -.0cm
\includegraphics[totalheight=0.27\textheight]{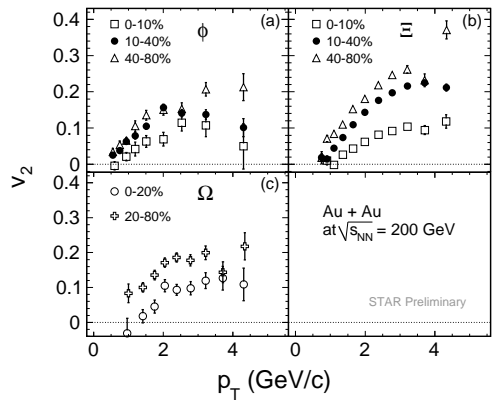}
\caption{The centrality dependence of $v_{2}$ as a function of $p_{T}$
in Au + Au collisions at \sqrtsNN = 200 GeV for (a) $\phi$, (b) $\Xi$ and (c) $\Omega$.
The error bars are shown only for the statistical uncertainties.} \label{fig1}
\end{figure*}
The centrality dependence of charged and strange hadron $v_{2}$ in Au + Au collisions
at \sqrtsNN = 200 GeV has been
well studied in Ref~\cite{flow3}. The large statistics from run 7 of
RHIC make the study on the centrality dependence of $v_{2}$ for multi-strange hadrons possible. The results are
shown in Fig. 1. Panel (a) and (b) show the $v_{2}$ as a function of
$p_{T}$ in $0-10\%$, $10-40\%$ and $40-80\%$ most central events for $\phi$
and $\Xi$. Due to the limited statistics, the $v_{2}$($p_{T}$) of $\Omega$ are shown
in two centrality bins, $0-20\%$ and $20-80\%$, in panel (c). The estimated
systematic error based on the background evaluation and track selection criteria
is around $10\%$. The larger $v_{2}$ values
could be observed in the more peripheral collisions. It is because the final anisotropy in
the momentum space is converted by the initial anisotropy of the collision geometry.
The larger eccentricity in the more peripheral collision drives the larger magnitude of $v_{2}$.

\begin{figure*}[ht]
\centering \hskip -.0cm \vskip -.0cm
\includegraphics[totalheight=0.27\textheight]{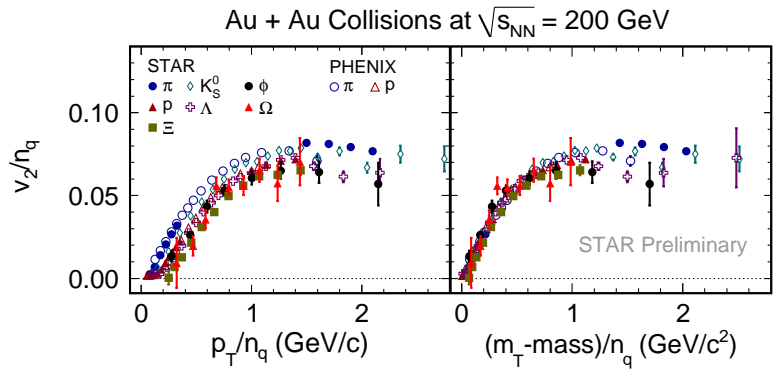}
\caption
{ The $v_{2}$ scaled by number of constituent quarks ($n_{q}$) as a function of
(a) $p_{T}/n_{q}$ and (b) $(m_{T} - \mathrm{mass})/n_{q}$ for identified particles in minimum bias
Au + Au collisions at \sqrtsNN = 200 GeV.
PHENIX data are from~\cite{PHENIX}. The error bars are shown only for the statistical uncertainties.
} \label{fig2}
\end{figure*}

The Number of Quark (NQ) scaling on $v_{2}$ in the intermediate $p_{T}$ range
(2 \GeVc $<p_{T}<$ 5 \GeVc) could be reproduced by the
quark coalescence~\cite{cola} or recombination~\cite{recom} mechanisms in
particle production. Thus, the NQ scaling indicates the deconfinement has been
achieved in the heavy ion collisions at RHIC. With the $v_{2}$ results from multi-strange
hadrons, we could test whether the scaling works for them. Multi-strange hadrons are regarded
as good probes to the early partonic stage of collision dynamics, because of their larger mass
and smaller hadronic cross section compared to the light-quark (u, d) hadrons.
Figure 2 shows the constituent-quark scaled \vtwo as a function of $p_{T}/n_{q}$ and
  $(m_{T} - \mathrm{mass})/n_{q}$ in panel (a) and (b), respectively.
It is known the NQ scaling works for identified charged hadrons ($\pi$, $K$ and $p$) and
strange hadrons (\ks and $\Lambda$)~\cite{flow4, starklv2}. The important information is
that multi-strange hadrons, especially $\phi$ and $\Omega$ which are pure $s$ constituent quark
contained hadrons, follow the NQ scaling up to $p_{T}/n_{q}\sim$1.5 \GeVc~or
$(m_{T} - \mathrm{mass})/n_{q}\sim$1 \GeVcsqr. This indicates that the major part of
the $v_{2}$ has been built up at the partonic stage. Hence,
 the partonic collectivity has been established at RHIC.

\begin{figure*}[ht]
\centering \hskip -.0cm \vskip -.0cm
\includegraphics[totalheight=0.31\textheight]{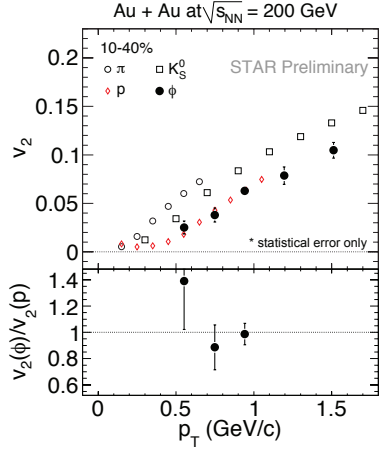}
\caption{
Upper panel: $v_{2}$ as a function of $p_{T}$ for $\pi$, ${K}^{0}_{S}$, $p$ and $\phi$ in
$10-40\%$ Au + Au collisions at \sqrtsNN = 200 GeV.
Lower panel: the ratio of $\phi$ $v_{2}$ to that of $p$.
The error bars are shown only for the statistical uncertainties.
} \label{fig3}
\end{figure*}

The mass ordering in data~\cite{flow3, flow4}, which is qualitatively consistent with ideal hydrodynamics,
works well for $\pi$, $K$, $p$, ${K}^{0}_{S}$, $\Lambda$ and $\Xi$.
Namely, the hadron with heavier mass shows the smaller $v_{2}$ in a given $p_{T}$ bin.
Recently, the calculations based on ideal hydrodynamical model
together with the hadron cascade~\cite{massordering} suggest that the mass ordering
of $v_{2}$ at low $p_{T}$ ($p_{T}$ $<$ 1.5 $\sim$ 2.0 \GeVc) could be
broken for $\phi$ mesons due to the small hadronic
cross section at late hadronic stage within their model.
%Here, $\phi$ mesons and proton are chosen to study
%the hadronic effect on $v_{2}$, because the mass of $\phi$ and proton are close, but
%$\phi$ mesons are less sensitive to hadronic rescatterings than protons. Thus protons
%would build up larger radial flow, which is the driving force of
%mass ordering~\cite{radialflow}, in the hadronic stage. It breaks the mass ordering for
%$\phi$ and proton in the low $p_{T}$ region.
Figure 3 shows the comparison in $10-40\%$ centrality bins. The mass ordering could be observed
for $\pi$, ${K}^{0}_{S}$ and $p$ clearly. For clarity, in the lower panel,
we show the ratio of $\phi$ meson $v_{2}$ to that of protons.
Within errors, they are consistent in the low $p_{T}$ region
($p_{T}<$ 1 \GeVc). Data with higher precision are needed for the final conclusion.

%\section{Summary}
In summary, we present the $v_{2}$ for multi-strange hadrons in \sqrtsNN = 200 GeV Au + Au collisions at STAR.
The centrality dependence of $v_{2}$ shows larger $v_{2}$ value in more peripheral collisions.
This is because of the larger initial anisotropy in the coordinate space.
The NQ scaling works for multi-strange hadrons in the intermediate $p_{T}$ range, it
indicates the partonic collectivity has been built up at RHIC.
In order to the study the late hadronic effect on $v_{2}$, the comparison of $\phi$ meson $v_{2}$ to
proton $v_{2}$ has been made in the low $p_{T}$ region. They are consistent within errors.
In the future, the performance of Time-Of-Flight detector in the tenth RHIC run at STAR will help us
on the further investigation.

\end{document}